\documentclass[12pt]{article}
\usepackage{epsf}
\usepackage{amsmath,amssymb}

\usepackage[dvips]{graphicx}

\usepackage{comment}

\begin{document}

\newcommand{\lsim}{\stackrel{<}{_\sim}}
\newcommand{\gsim}{\stackrel{>}{_\sim}}

\newcommand{\rem}[1]{{$\spadesuit$\bf #1$\spadesuit$}}

\renewcommand{\thefootnote}{\fnsymbol{footnote}}
\setcounter{footnote}{0}

\begin{titlepage}

\def\thefootnote{\fnsymbol{footnote}}

\begin{center}

\hfill UT-16-01\\
\hfill January, 2016\\

\vskip .75in

{\Large \bf 

  Studying 750 GeV Di-photon Resonance
  \\
  at Photon-Photon Collider
  \\

}

\vskip .5in

{\large 
  Hayato Ito, Takeo Moroi and Yoshitaro Takaesu
}

\vskip 0.25in

\vskip 0.25in

{\em Department of Physics, University of Tokyo,
Tokyo 113-0033, Japan}

\end{center}
\vskip .5in

\begin{abstract}

  Motivated by the recent LHC discovery of the di-photon excess at the
  invariant mass of $\sim 750\ {\rm GeV}$, we study the prospect of
  investigating the scalar resonance at a future photon-photon collider.  We
  show that, if the di-photon excess observed at the LHC is due to a
  new scalar boson coupled to the standard-model gauge bosons, such a
  scalar boson can be observed and studied at the photon-photon
  collider with the center-of-mass energy of $\sim 1\ {\rm TeV}$ in
  large fraction of parameter space.
  
\end{abstract}

\end{titlepage}

\renewcommand{\thepage}{\arabic{page}}
\setcounter{page}{1}
\renewcommand{\thefootnote}{\#\arabic{footnote}}
\setcounter{footnote}{0}

Recently, both ATLAS and CMS collaborations have reported the results
of their analysis of di-photon events.  The result of ATLAS, which is
based on the 2015 data with the integrated luminosity of $3.2\ {\rm
  fb}^{-1}$, shows the excess with the local (global) significance of
3.6$\sigma$ (2.0$\sigma$) at the di-photon invariant-mass of
$M_{\gamma\gamma}\sim 750\ {\rm GeV}$ \cite{ATLAS-CONF-2015-081}.  In
addition, the CMS result, which is based on their 2015 data with the
integrated luminosity of $2.6\ {\rm fb}^{-1}$, also indicates an
excess at similar $M_{\gamma\gamma}$; the local significance of the
excess is claimed to be 2.6$\sigma$ (while the global significance
is smaller than 1.2$\sigma$) \cite{CMS-PAS-EXO-15-004}.  These
excesses may indicate the existence of a new resonance $\Phi$ with its
mass of $m_\Phi\sim 750\ {\rm GeV}$.  The production cross section and
branching ratio into di-photon are suggested to satisfy
\cite{Nakai:2015ptz}--\cite{Ellis:2015oso}
\begin{align}
  \sigma (pp\rightarrow \Phi \rightarrow \gamma\gamma)
  = \sigma (pp\rightarrow \Phi) {\rm Br}(\Phi \rightarrow\gamma\gamma)
  \sim 5-10\ {\rm fb}:
  ~ \mbox{at} ~\sqrt{s_{pp}}=13\ {\rm TeV},
\end{align}
where $\sqrt{s_{pp}}$ is the center-of-mass energy of $pp$.

One of the explanations of such an excess is a new scalar boson
 coupled to standard-model gauge bosons via higher dimensional
 operators~\cite{Nakai:2015ptz}--\cite{Deppisch:2016scs}.
The scalar boson may interact as
\begin{align}
  {\cal L}_{\rm eff} = 
  \frac{1}{2\Lambda_i} \varphi
  \epsilon^{\mu\nu\rho\sigma}
  {\cal F}_{\mu\nu}^{(i),a} {\cal F}_{\rho\sigma}^{(i),a},
  \label{Lvarphi}
\end{align}
for the case of pseudo-scalar boson $\varphi$, or as
\begin{align}
  {\cal L}_{\rm eff} = 
  \frac{1}{\Lambda_i} \phi
  {\cal F}_{\mu\nu}^{(i),a} {\cal F}_{\mu\nu}^{(i),a},
  \label{Lphi}
\end{align}
for the case of real scalar boson $\phi$.  Here, ${\cal
  F}_{\mu\nu}^{(i),a}$ is the field strength of the standard-model
gauge bosons where $i=1$, $2$, and $3$ correspond to $U(1)_Y$,
$SU(2)_L$, and $SU(3)_C$, respectively, and the superscript $a$
denotes the index for the adjoint representations.  The summations
over the repeated indices are implicit.  With these interactions, the
scalar boson can be produced at the LHC via the gluon-gluon scattering
process, and it can decay into photon pair.  Then, the di-photon excess
observed at the LHC can be explained by the process of $gg\rightarrow
\Phi\rightarrow\gamma\gamma$ (where $\Phi=\phi$ or $\varphi$, and $g$
is gluon).

If there exists such a new scalar boson, the next task for future
collider experiments is to precisely study its properties.  
The LHC will play very important role in such a
program~\cite{Alves:2015jgx},\cite{Cao:2015pto}--\cite{Csaki:2016raa}.  
Here, we
consider another possibility, which is the photon-photon collider.
The photon-photon collider can be realized at the facility of the
International $e^+e^-$ Linear Collider (ILC) by converting electron
beam to high energy photons.  Because the scalar field of our interest
decays into di-photon, it can be produced via the photon-photon
scattering process.  For the case of the scalar fields with the
interactions given in Eq.\ \eqref{Lvarphi} or \eqref{Lphi}, one of the
advantages of the photon-photon collider is that the single production
of the scalar boson is possible so that the kinematical reach is close
to the total center-of-mass energy of the collider.  In the case of
our interest where the mass of the scalar boson is about $m_\Phi\sim
750\ {\rm GeV}$, the scalar boson can be produced at the photon-photon
collider with $\sqrt{s_{ee}}\sim 1\ {\rm TeV}$ (where
$\frac{1}{2}\sqrt{s_{ee}}$ is the energy of the initial-state electron
beam), which is within the range of the upgrade ILC.  In addition,
the clean environment of the photon-photon collider may help precisely
studying the properties of the scalar boson.

In this letter, we study the prospect for the study of the scalar
boson at the photon-photon collider.  Assuming the effective
Lagrangian given above, we calculate the production cross section of
the scalar boson.  Because the effective interaction may originate
from the loop effects of heavy vector-like particles or from unknown
strong dynamics, we do not specify the origin of the effective
Lagrangian and treat $\Lambda_i$ as free parameters.  We also estimate
the cross sections for standard-model backgrounds.  We will see that,
if the di-photon excess observed at the LHC is due to a scalar boson
with its mass of $m_\Phi\sim 750\ {\rm GeV}$ and couplings given in
Eq.\ \eqref{Lvarphi} or \eqref{Lphi}, such a new scalar boson can be
observed and studied at the photon-photon collider in large fraction
of the paramter space.  In particular, the cross section for the
process $\gamma \gamma \rightarrow \Phi \rightarrow gg$ is large
enough so that such a process can be observed at the photon-photon
collider.  In addition, some of the processes $\gamma \gamma
\rightarrow \Phi \rightarrow\gamma\gamma$, $\gamma Z$, $ZZ$, or
$W^+W^-$ may be also observed.  With these observations, information
about the parameters $\Lambda_i$ in Eq.\ \eqref{Lvarphi} or
\eqref{Lphi} is obtained.

First, we summarize the formulas of the cross sections for the scalar
production processes at the LHC and ILC.  Hereafter, we concentrate on
the case where $\Gamma_\Phi\ll m_\Phi$ (with $\Gamma_\Phi$ being the
total decay width of $\Phi$).  Then, assuming that the gluon-gluon
scattering process is the dominant production process at the LHC, the
cross section for the processes
$pp\rightarrow\Phi\rightarrow\gamma\gamma$ is given by
\begin{align}
  \sigma (pp\rightarrow \Phi\rightarrow \gamma\gamma) =
  \frac{\pi^2}{8 m_\Phi s_{pp}}
  \frac
  {\Gamma (\Phi\rightarrow gg) \Gamma (\Phi\rightarrow \gamma\gamma)}
  {\Gamma_\Phi}
  \int dx_1 dx_2 \delta (x_1 x_2 - m_\Phi^2/s_{pp}) g(x_1) g(x_2),
  \label{sigma(LHC)}
\end{align}
where $g(x)$ is the parton distribution function (PDF) of gluon.  We
use MSTW2008NLO PDF \cite{Martin:2009iq}, with which the integral in Eq.\
\eqref{sigma(LHC)} is evaluated to be $1.7\times 10^3$ for
$m_\Phi=750\ {\rm GeV}$ and $\sqrt{s_{pp}}=13\ {\rm TeV}$.  In
addition, $\Gamma (\Phi\rightarrow gg)$ and $\Gamma (\Phi\rightarrow
\gamma\gamma)$ are partial decay widths for the $gg$ and
$\gamma\gamma$ final states, respectively.

For the monochromatic photon beams, the cross section for the process
$\gamma\gamma\rightarrow \Phi\rightarrow F$ at
$\sqrt{s_{\gamma\gamma}}\sim m_\Phi$ (with $\sqrt{s_{\gamma\gamma}}$
being the center-of-mass energy of $\gamma\gamma$) is given by
\begin{align}
  \hat{\sigma}
  (\gamma\gamma\rightarrow \Phi \rightarrow F; s_{\gamma\gamma}) =
  \left( \frac{1+\xi_2\xi_2'}{2} \right) \times
  \frac{16\pi m_\Phi^2}{s_{\gamma\gamma}}
  \frac{\Gamma (\Phi \rightarrow \gamma \gamma)
    \Gamma (\Phi \rightarrow F)}
  {(s_{\gamma\gamma}-m_\Phi^2)^2 + m_\Phi^2 \Gamma_\Phi^2},
  \label{sighat}
\end{align}
where $\xi_2$ and $\xi'_2$ are Stokes parameters of the initial-state
photons, where $\xi_2=\pm 1$ corresponds to the photons with helicity
$\pm 1$.

At the photon-photon collider, the initial-state photons are provided
as back-scattered photons off the electron beams, and are not
monochromatic.  For the calculation of the cross section at the
photon-photon collider, we use the luminosity function of the
back-scattered photons given in~\cite{Ginzburg:1982yr, Ginzburg:1999wz}:
\begin{align}
\frac{1}{L_{ee}} \frac{d^2L_{\gamma\gamma}}{dy dy'} = 
  f(x, y) B(x, y) f(x, y') B(x, y'),
\end{align}
where $L_{ee}$ is the luminosity of the electron beam, $y$ and $y'$ denote the photon energies normalized by the energy
of the electron beam $\frac{1}{2}\sqrt{s_{ee}}$, and $x\equiv
2\sqrt{s_{ee}}\omega_0/m_e^2$, with $m_e$ being the electron mass and
$\omega_0$ the averaged energy of the laser photons in a laboratory
frame.  The function $f$ is given in the following form:
\begin{align}
  f(x, y) = \frac{2\pi\alpha^2}{\sigma_c x m_e^2} C_{00}(x,y),
\end{align}
where $\alpha$ is the QED fine structure constant, and
\begin{align}
  C_{00}(x,y) = \frac{1}{1-y} - y +( 2r-1 )^2
  - \lambda_e P_l x r (2r-1) (2-y),
  \label{C00}
\end{align}
with $r\equiv y /x (1-y)$.  In Eq.\ \eqref{C00}, $\lambda_e/2$ and
$P_l$ are the mean helicities of initial electrons and laser photons,
respectively.  In our numerical calculation, we take $\lambda_e=0.85$
and $P_l=-1$.  In addition,
\begin{align}
  \sigma_c = \sigma_c^{\rm (np)} + \lambda_e P_l \sigma_1,
\end{align}
where 
\begin{align}
  \sigma_c^{\rm (np)} &= \frac{2\pi \alpha^2}{x m^2_e}
  \left[
    \left(
      1 - \frac{4}{x} - \frac{8}{x^2}
    \right)
    \ln(x+1) + \frac{1}{2} + \frac{8}{x} - \frac{1}{2 (x+1)^2}
  \right],
  \\
  \sigma_1 &= \frac{2\pi \alpha^2}{x m^2_e}
  \left[ 
    \left(
      1 + \frac{2}{x}
    \right)
    \ln(x+1) - \frac{5}{2} + \frac{1}{x+1} - \frac{1}{2(x+1)^2}
  \right].
\end{align}
The function $B$ is given by
\begin{align}
  B (x,y) = 
  \left\{ 
    \begin{array}{ll}
      \displaystyle{
        \exp 
        \left[ -\frac{\rho^2}{8} \left( \frac{x}{y} - x - 1 \right) \right]
      }
      & ~:~ y_m/2 < y < y_m
      \\
      0 & ~:~ \mbox{otherwise}
    \end{array}
  \right. ,
\end{align}
where $y_m\equiv x/(x+1)$, and $\rho$ is the reduced distance between
conversion and collision points.  We assume the photon-photon collider
with $x = 4.8$, which maximizes $y_m$ without spoiling the photon
luminosity \cite{Ginzburg:1982yr, Ginzburg:1999wz}, and take $\rho=1$
in our numerical calculation.  In addition, the Stokes parameter is
given by
\begin{align}
  \xi_2(y) = \frac{C_{20}(x,y)}{C_{00}(x,y)},
\end{align}
where
\begin{align}
  C_{20}(x,y) = \lambda_e r x 
  \left[ 1 + ( 1 - y )( 2r -1 )^2 \right]
  -P_l ( 2 r - 1 ) \left( \frac{1}{1-y} + 1-y \right).
  \label{C20}
\end{align}
Notice that $\xi_2(y)\rightarrow 1$ as $y\rightarrow y_m$.
Using the luminosity function of the back-scattered photons, the cross
section at the photon-photon collider is given by
\begin{align}
  \sigma
  (\gamma\gamma\rightarrow \Phi \rightarrow F) 
  &= 
  \frac{1}{L_{ee}} 
  \int_0^{y_m} dy dy'
  \frac{d^2L_{\gamma\gamma}}{dy dy'}
  \hat{\sigma}
  (\gamma\gamma\rightarrow \Phi \rightarrow F; 
  s_{\gamma\gamma} = y y' s_{ee}).
  \label{sigma(photonphoton)}
\end{align}
We normalize
the cross section at the photon-photon collider using $L_{ee}$.

For the calculation of the cross sections, it is important to
understand the decay widths of the scalar boson.  In the following
numerical analysis, we consider the case where the scalar resonance is
a pseudo-scalar boson $\varphi$.  (For $m_\Phi\gg m_Z$, the following
results are almost the same even if we consider the case of real
scalar boson.)  In such a case, adopting the effective Lagrangian
given in Eq.\ \eqref{Lvarphi}, the partial decay widths of $\varphi$
are given by
\begin{align}
  \Gamma (\varphi\rightarrow gg) &= 
  \frac{2 m_\varphi^3}{\pi \Lambda_3^2},
  \\
  \Gamma (\varphi\rightarrow \gamma\gamma) &= 
  \frac{m_\varphi^3}{4 \pi \Lambda_{\gamma\gamma}^2},
  \\
  \Gamma (\varphi\rightarrow \gamma Z) &= 
  \frac{m_\varphi^3}{8 \pi \Lambda_{\gamma Z}^2}
  \left( 1 - \frac{m_Z^2}{m_\varphi^2} \right)^3,
  \\
  \Gamma (\varphi\rightarrow Z Z) &= 
  \frac{m_\varphi^3}{4 \pi \Lambda_{ZZ}^2}
  \left( 1 - \frac{4m_Z^2}{m_\varphi^2} \right)^{3/2},
  \\
  \Gamma (\varphi\rightarrow W^+ W^-) &= 
  \frac{m_\varphi^3}{2 \pi \Lambda_2^2}
  \left( 1 - \frac{4m_W^2}{m_\varphi^2} \right)^{3/2},
\end{align}
where $m_\varphi$, $m_Z$, and $m_W$ are the masses of pseudo-scalar
boson $\varphi$, $Z$ boson, and $W^\pm$ boson, respectively, and
\begin{align}
  \Lambda_{\gamma\gamma}^{-1} &\equiv
  \Lambda_2^{-1} \sin^2 \theta_W + \Lambda_1^{-1} \cos^2 \theta_W,
  \\
  \Lambda_{\gamma Z}^{-1} &\equiv
  2 (\Lambda_2^{-1} - \Lambda_1^{-1}) \sin\theta_W \cos \theta_W,
  \label{Lambda_gammaZ}
  \\
  \Lambda_{ZZ}^{-1} &\equiv 
  \Lambda_2^{-1} \cos^2 \theta_W + \Lambda_1^{-1} \sin^2 \theta_W,
\end{align}
with $\theta_W$ being the Weinberg angle.  (The decay widths of the
real scalar boson can be found, for example, in \cite{Alves:2015jgx}.)
In our analysis, the total decay width is treated as a free parameter,
because there may exist decay modes other than $\Phi\rightarrow VV'$
(with $V$ and $V'$ denoting standard-model gauge bosons), like the
decay into dark matter pairs \cite{Backovic:2015fnp,Bi:2015uqd,Bauer:2015boy,Dev:2015isx,Davoudiasl:2015cuo,Park:2015ysf,Mambrini:2015wyu,Han:2015yjk}.  Thus, there are
four free parameters in our analysis, i.e., $\Lambda_1$, $\Lambda_2$,
$\Lambda_3$, and $\Gamma_\Phi$.  With $\sigma
(pp\rightarrow\Phi\rightarrow\gamma\gamma)$ being fixed, the cross
section $\sigma (\gamma\gamma\rightarrow\Phi\rightarrow VV')$, which
is our primary interest, is insensitive $\Gamma_\Phi$ in particular
when $\Gamma_\Phi \ll m_\Phi$, as we discuss in the following.

Using Eq.\ \eqref{sigma(photonphoton)}, we calculate the total cross
section for the scalar-boson production process at the photon-photon
collider.  Assuming that the $\Phi$ production at the LHC is via the
gluon-fusion process, the cross sections
$\sigma (\gamma \gamma \rightarrow \Phi\rightarrow F)$ and
$\sigma (pp\rightarrow \Phi\rightarrow\gamma\gamma)$ are related as
\begin{align}
  \sigma (\gamma \gamma \rightarrow \Phi\rightarrow F)
  \simeq \sigma_{gg}
  \times
  \frac{{\rm Br}(\Phi\rightarrow F)}{{\rm Br}(\Phi\rightarrow gg)}
  \times
  \left[
    \frac{\sigma (pp\rightarrow \Phi\rightarrow\gamma\gamma;
      \sqrt{s_{pp}}=13\ {\rm TeV})}
    {10\ {\rm fb}}
  \right],
  \label{sigma_fitting}
\end{align}
where $\sigma_{gg}$ is a numerical constant.  (See Eqs.\
\eqref{sigma(LHC)}, \eqref{sighat} and \eqref{sigma(photonphoton)}.)
In our numerical calculation, we take $\sigma (pp\rightarrow
\Phi\rightarrow\gamma\gamma; \sqrt{s_{pp}}=13\ {\rm TeV})=10\ {\rm fb}$.  
We note here that, although we mostly consider the decay of $\Phi$
into gauge-boson pairs, the above equation is applicable to any final
state.

We calculate $\sigma_{gg}$ for the narrow width case (i.e.,
$\Gamma_\Phi\ll m_\Phi$) as well as for the cases of $\Gamma_\Phi=25$
and $50\ {\rm GeV}$.  For the latter cases, the integration in Eq.\
\eqref{sigma(photonphoton)} is performed in the region of
$m_\Phi-2\Gamma_\Phi\leq\sqrt{s_{\gamma\gamma}}\leq
m_\Phi+2\Gamma_\Phi$, because the photon luminosity function we adopt
is reliable only around $\sqrt{s_{\gamma\gamma}}\sim m_\Phi$; we have
checked that the result does not change much even if we perform the
integration in the region of
$m_\Phi-\Gamma_\Phi\leq\sqrt{s_{\gamma\gamma}}\leq
m_\Phi+\Gamma_\Phi$.

\begin{figure}[t]
  \centerline{\epsfxsize=0.55\textwidth\epsfbox{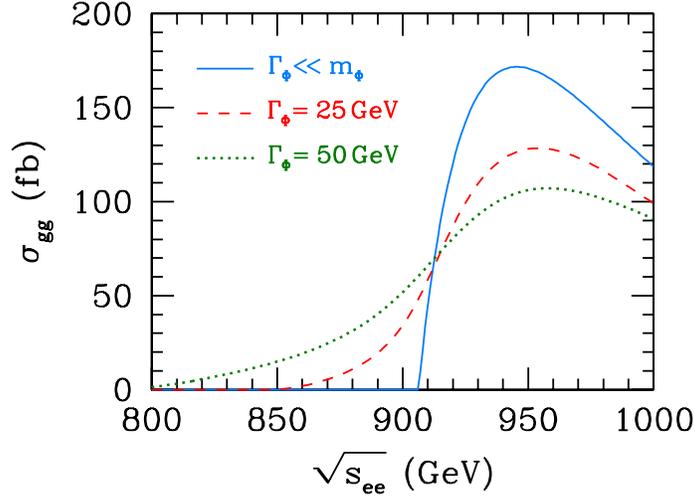}}
  \caption{\small $\sigma_{gg}$ as a function of $\sqrt{s_{ee}}$ for
    $m_\Phi=750\ {\rm GeV}$. The blue (solid) line is for the narrow
    width case, while the red (dashed) and green (dotted) ones are for
    $\Gamma_\Phi=25$ and $50\ {\rm GeV}$, respectively.}
  \label{fig:sigma0}
\end{figure}

In Fig.\ \ref{fig:sigma0}, we plot $\sigma_{gg}$ as a function of
$\sqrt{s_{ee}}$, taking $m_\Phi=750\ {\rm GeV}$.  We can see that
$\sigma_{gg}$ is maximized when $\sqrt{s_{ee}}\sim 950\ {\rm GeV}$.
The maximal value of $\sigma_{gg}$ depends on the width of the scalar
boson, and $\sigma_{gg}$ can be as large as $110-170\ {\rm fb}$ for
$\Gamma_\Phi\lesssim 50\ {\rm GeV}$.  We are using the photon
luminosity function which has a peak at $\sqrt{s_{\gamma\gamma}}\simeq
0.79\sqrt{s_{ee}}$.  Thus, in the narrow width case, $\sigma_{gg}$
takes its maximal value of $170\ {\rm fb}$ for $\sqrt{s_{ee}}\simeq
945\ {\rm GeV}$.  With the integrated luminosity of $L_{ee}\sim O(1)\
{\rm ab}^{-1}$, the total number of the events at the photon-photon
collider is as large as (or larger than) $O(10^5)$, which may make the
detailed study of the scalar boson $\Phi$ possible at the
photon-photon collider.  (We note that we neglect the interference
between the scalar-boson-exchange and standard-model diagrams, and
hence our results are not reliable for the parameter region where the
signal cross sections become comparable to the background ones.)

In order to discuss the collider phenomenology of the scalar boson at
the photon-photon collider, we need to understand background
processes.  Assuming the effective Lagrangian given in
Eq.\ \eqref{Lvarphi} or \eqref{Lphi}, we consider the signal processes
of
\begin{align*}
  \gamma \gamma \rightarrow \Phi \rightarrow VV',
\end{align*}
where $VV'=\gamma\gamma$, $\gamma Z$, $ZZ$, $W^+ W^-$, and $gg$.  For
these signal events, there exist standard-model backgrounds.  In order
to estimate the number of backgrounds for $VV'=\gamma\gamma$, $\gamma
Z$, $ZZ$, and $W^+ W^-$, we calculate the following quantity in the
standard-model:
\begin{align}
  \kappa_{VV'} \equiv
  \left.
    \frac{d\sigma^{\rm (SM)} (\gamma \gamma \rightarrow VV')}{d m_{VV'}}
  \right|_{|\cos\theta_{\rm CM}|<\cos\theta_{\rm cut},\ m_{VV'}=m_\Phi},
\end{align}
where $m_{VV'}$ is the invariant mass of the final-state gauge boson
pair.  Some of the final-state gauge bosons are likely to be emitted
to the beam direction.  Thus, in order to reduce the number of
backgrounds, the phase-space integration is limited to the region of
$|\cos\theta_{\rm CM}|<\cos\theta_{\rm cut}$, where $\theta_{\rm CM}$
is the angle between initial-state photon and the final-state gauge
boson in the center-of-mass frame.  Correspondingly, in the
calculation of the signal cross section, the phase-space integration
should be performed in the region of $|\cos\theta_{\rm
  CM}|<\cos\theta_{\rm cut}$.  Because the final-state particles are
emitted spherically in the signal events, the effect of limiting the
final-state phase space is taken into account by multiplying the total
cross section by $\cos\theta_{\rm cut}$.  We expect to reduce the
standard-model backgrounds by using the kinematical variable $m_{VV'}$
(as well as $\cos\theta_{\rm CM}$), and hence the expected numbers of
the background events are estimated as
\begin{align}
  N_{VV'}^{\rm (BG)} = L_{ee} \kappa_{VV'} \Delta m_{VV'},
\end{align}
where $\Delta m_{VV'}$ is the width of the bin for the study of the
scalar boson production.  In the standard model, the processes
$\gamma\gamma\rightarrow\gamma\gamma$, $\gamma Z$, and $ZZ$ happen at
the loop level, and their cross sections can be found in
\cite{Gounaris:1999gh, Gounaris:1999ux, Gounaris:1999hb}.  On the contrary,
the process $\gamma\gamma\rightarrow W^+ W^-$ occurs at the tree
level.  The cross section for this process is evaluated with the use
of {\tt MadGraph5} \cite{Alwall:2014hca}.  The values of
$\kappa_{VV'}$ (with $VV'=\gamma\gamma$, $\gamma Z$, $ZZ$, and $W^+
W^-$) are summarized in Table \ref{table:bgVV}.

\begin{table}[t]
  \begin{center}
    \begin{tabular}{cccccc}
      \hline\hline
      {$VV'$}
      & {$\kappa_{VV'} (0.99) $}
      & {$\kappa_{VV'} (0.95) $}
      & {$\kappa_{VV'} (0.90) $}
      & {$\kappa_{VV'} (0.80) $}
      & {$\kappa_{VV'} (0.70) $} \\
      \hline
      {$\gamma\gamma$} & {$0.057$} & {$0.049$} & {$0.041$}& {$0.032$}& {$0.025$} \\
      {$\gamma Z$} & {$0.36$} &{$0.30$} & {$0.26$}& {$0.20$}& {$0.16$} \\
      {$ZZ$} &{$0.58$} & {$0.50$} & {$0.42$} & {$0.32$}& {$0.25$}\\
      {$W^+W^-$} &{$137$} & {$67$} & {$42$} & {$24$}& {$16$}\\
      \hline\hline
    \end{tabular}
    \caption{\small The value of $\kappa_{VV'}(\cos\theta_{\rm cut})$
      with $VV'=\gamma\gamma$, $\gamma Z$, $ZZ$, and $W^+ W^-$ in
      units of ${\rm fb/GeV}$.  Here, we take $\sqrt{s_{ee}}=945\ {\rm
        GeV}$.}
    \label{table:bgVV}
\vspace{1cm}
    \begin{tabular}{cccccc}
      \hline\hline
      {$m_{\bar{q}q}$}
      & {$\kappa_{\bar{q}q} (0.99) $}
      & {$\kappa_{\bar{q}q} (0.95) $}
      & {$\kappa_{\bar{q}q} (0.90) $}
      & {$\kappa_{\bar{q}q} (0.80) $}
      & {$\kappa_{\bar{q}q} (0.70) $} \\
      \hline
      {$750\ {\rm GeV}$} &{$0.44$} & {$0.28$} & {$0.21$}& {$0.14$}& {$0.11$}\\
      {$725\ {\rm GeV}$} &{$1.0$} & {$0.62$} & {$0.47$}& {$0.33$}& {$0.24$}\\
      {$700\ {\rm GeV}$} &{$1.6$} & {$1.0$} & {$0.75$}& {$0.52$}& {$0.38$}\\
      \hline\hline
    \end{tabular}
    \caption{\small The value of $\kappa_{\bar{q}q}(\cos\theta_{\rm
        cut})$ in units of ${\rm fb/GeV}$ for several values of
      $m_{\bar{q}q}$.  Here, we take $\sqrt{s_{ee}}=945\ {\rm GeV}$.}
    \label{table:bgqq}
  \end{center}
\end{table}

For the gluon-gluon final state, the dominant background is from the
pair-production of light quarks, $\gamma\gamma\rightarrow \bar{q}q$,
because gluon and light-quark jets are indistinguishable.  Thus, we
calculate
\begin{align}
  \kappa_{\bar{q}q} \equiv
  \sum_{q=u,d,s,c}
  \left.
    \frac{d\sigma^{\rm (SM)} (\gamma \gamma \rightarrow \bar{q}q)}
    {d m_{\bar{q}q}}
  \right|_{|\cos\theta_{\rm CM}|<\cos\theta_{\rm cut}},
  \label{kappa_qq}
\end{align}
where $m_{\bar{q}q}$ is the invariant mass of the final-state
$\bar{q}q$ system, and $\cos\theta_{\rm CM}$ denotes the angle between
the initial-state photon and the final-state quark in the
center-of-mass frame.  (For our numerical calculation, {\tt MadGraph5}
is used.)  Because we expect a high $b$-tagging efficiency at the ILC
detectors, i.e., $\epsilon_b\sim 70-80\ \%$ with the
mis-identification efficiency of light quarks being $O(1)\ \%$
\cite{Behnke:2013lya}, we assume that the $\bar{b}b$ contribution can
be largely eliminated and hence is sub-dominant.  Thus, we do not
include its effect into the calculation of $\kappa_{\bar{q}q}$.  We
found that $\kappa_{\bar{q}q}$ has a sizable dependence on
$m_{\bar{q}q}$, and hence we calculate this quantity for several
values of $m_{\bar{q}q}$.  The results are shown in Table
\ref{table:bgqq}.  Importantly, the light-quark pair production cross
section is strongly dependent on the poralization of initial-state
photons.  In particular, the cross section vanishes when the stokes
parameters of two initial-state photons satisfy $\xi_2\xi_2'=1$; with
our choice of collider parameters, this relation is realized at the
end-point of the spectrum of back-scattered photon, i.e., $y=y'=y_m$.
Even at the peak of the photon luminosity function, the product
$\xi_2\xi_2'$ is close to $1$, and the light-quark pair production
cross section is significantly suppressed compared to the case of
unpolarized initial-state photons.

Now we discuss the signal-to-background ratio for various final
states.  For this purpose, we specify relevant size of the bin,
$\Delta m_{VV'}$, for the study of the scalar-boson production.  If
the width of the scalar boson is narrow enough, $\Delta m_{VV'}$ is
determined by the detector resolution (where we consider the events
with no missing momentum).  It is expected that, at the ILC detectors,
the energy of charged particles and hadronic jets will be measured
with excellent accuracies \cite{Behnke:2013lya}.  With the ILC
detectors, the jet energy will be measured with the accuracy of
$3\ \%$ or better for jet energies above $100\ {\rm GeV}$.  In
addition, the energy resolution of the electromagnetic calorimeter of
the SiD detector, for example, is expected to be $\delta
E/E=0.17/\sqrt{E}\oplus 1\ \%$ for electrons or photons, and hence we
expect that the energy of the photons emitted by the decay of $\Phi$
will be measured with the uncertainty of $\sim 1\ \%$.  

Combining Eq.\ \eqref{sigma_fitting} and Fig.\ \ref{fig:sigma0}, the
cross section for the gluon-gluon final state at the photon-photon
collider is estimated to be $170\ {\rm fb}$ in the narrow width case,
taking $\sigma (pp\rightarrow \Phi\rightarrow\gamma\gamma)= 10\ {\rm
  fb}$ and $\sqrt{s_{ee}}=945\ {\rm GeV}$.  (The cross section becomes
smaller as $\Gamma_\Phi$ increases.)  On the contrary, using the
results given in Table \ref{table:bgqq}, the cross section of the
$\bar{q}q$ background is significantly smaller than the signal cross
section even if we take $\Delta m_{VV'}\sim O(10)\ {\rm GeV}$.
Integrating out $\kappa_{\bar{q}q}$ for $0.94\,m_\Phi\leq
m_{\bar{q}q}\leq 1.06\,m_\Phi$, for example, we found that the
background cross section is $23\ {\rm fb}$ for $\cos\theta_{\rm cut}
=0.9$.  Thus, it will be possible to observe and study signal events
with gluon-gluon final state with the luminosity of $L_{ee}\sim
O(0.1-1)\ {\rm fb}^{-1}$ or larger.

\begin{figure}[t]
  \centerline{\epsfxsize=0.55\textwidth\epsfbox{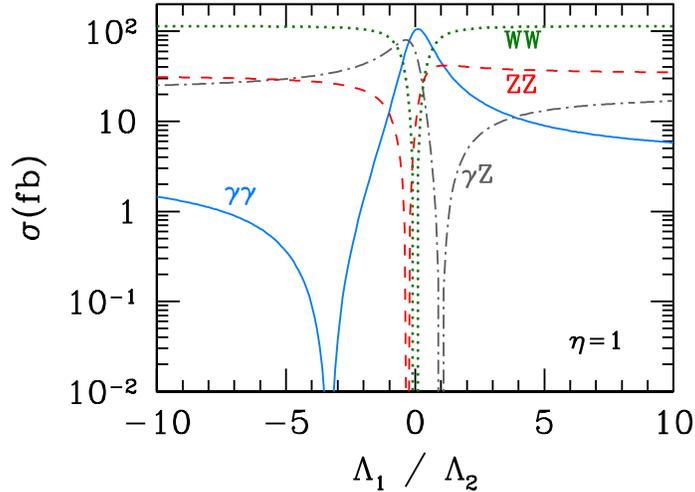}}
  \caption{\small Cross sections for the processes of
    $\gamma\gamma\rightarrow\Phi\rightarrow\gamma\gamma$ (blue solid),
    $\gamma Z$ (gray dot-dashed), $ZZ$ (red dashed), and $W^+ W^-$
    (green dotted) as functions of the ratio $\Lambda_1/\Lambda_2$.
    Here, we take $\sqrt{s_{ee}}=945\ {\rm GeV}$, $\eta =1$, and
    $\sigma (pp\rightarrow \Phi\rightarrow\gamma\gamma;
    \sqrt{s_{pp}}=13\ {\rm TeV})=10\ {\rm fb}$.}
  \label{fig:csvv}
\end{figure}

For the final states containing weak bosons and photon, with
$\Gamma_\Phi$ being fixed, the cross sections depend on $\Lambda_3$
through the following ratio:
\begin{align}
  \eta \equiv
  \frac{\Gamma (\Phi\rightarrow\gamma\gamma) + 
    \Gamma (\Phi\rightarrow\gamma Z) + \Gamma (\Phi\rightarrow Z Z) + 
    \Gamma (\Phi\rightarrow W^+ W^-)}{\Gamma (\Phi\rightarrow gg)}.
\end{align}
In the limit of $m_\Phi\gg m_Z$, $\eta\simeq
(\frac{1}{8}\Lambda_1^{-2} +\frac{3}{8}\Lambda_2^{-2})
\Lambda_3^2$. Then, the cross sections for the processes
$\gamma\gamma\rightarrow\Phi\rightarrow\gamma\gamma$, $\gamma Z$,
$ZZ$, and $W^+ W^-$ depend on two parameters, i.e., $\eta$ and the
ratio $\Lambda_1/\Lambda_2$ (as far as $\Gamma_\Phi$ and $\sigma
(pp\rightarrow \Phi\rightarrow\gamma\gamma)$ are fixed).  In addition,
the cross sections for these processes are proportional to $\eta$ for
a given value of $\Lambda_1/\Lambda_2$.  In Fig.\ \ref{fig:csvv},
taking $\eta=1$, we plot the cross sections of these processes as
functions of $\Lambda_1/\Lambda_2$.  We can see that $\sigma
(\gamma\gamma\rightarrow\Phi\rightarrow\gamma Z)$ becomes suppressed
when $\Lambda_1\simeq\Lambda_2$; this is because, in such a case, the
$\Phi$-$\gamma$-$Z$ coupling is suppressed (see Eq.\
\eqref{Lambda_gammaZ}).  In addition, the cross sections $\sigma
(\gamma\gamma\rightarrow\Phi\rightarrow\gamma\gamma)$ and $\sigma
(\gamma\gamma\rightarrow\Phi\rightarrow ZZ)$ are suppressed when
$\Lambda_1\simeq -\cot^2\theta_W\Lambda_2$ and $\Lambda_1\simeq
-\tan^2\theta_W\Lambda_2$, respectively.  We note here that the
present scenario is constrained by the searches for the resonances
which decay into a pair of electroweak gauge bosons performed at the
LHC Run-1.  In particular, the negative search for the resonance which
decays into $\gamma Z$ final state gives the most stringent
constraint.  According to \cite{Franceschini:2015kwy}, for example,
the parameter region of $\Gamma (\Phi\rightarrow\gamma Z)\gtrsim
2\Gamma (\Phi\rightarrow\gamma \gamma)$ is excluded, where we adopted
$\sigma (pp\rightarrow \Phi \rightarrow \gamma\gamma)\sim 10\ {\rm
  fb}$.  The regions of $\Lambda_1/\Lambda_2\lesssim -1$ and
$\Lambda_1/\Lambda_2\gtrsim 6$ conflict with such a constraint.  (If
we adopt a smaller value of $\sigma (pp\rightarrow \Phi \rightarrow
\gamma\gamma)$, the constraint becomes weaker.)

\begin{figure}[t]
  \centerline{\epsfxsize=0.55\textwidth\epsfbox{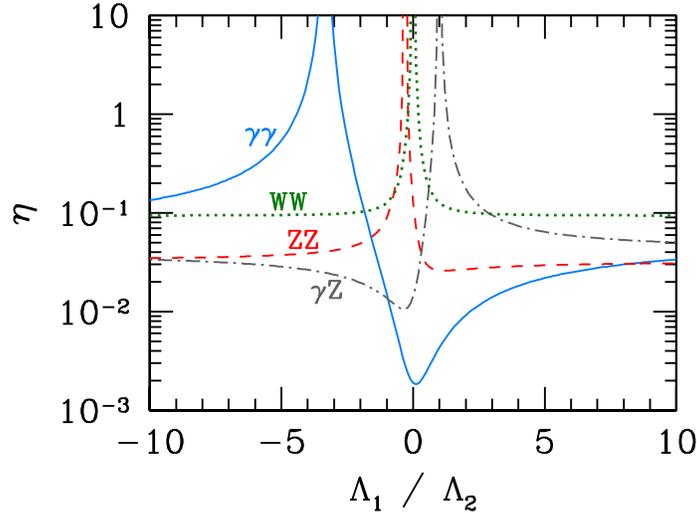}}
  \caption{\small The minimal values of $\eta$ to realize $S_{VV'} /
    \sqrt{N_{VV'}^{\rm (BG)}}>5$ for $VV'=\gamma\gamma$ (blue solid),
    $\gamma Z$ (gray dot-dashed), $ZZ$ (red dashed), and $W^+ W^-$
    (green dotted) as functions of the ratio $\Lambda_1/\Lambda_2$.
    The size of the bin is taken to be $\Delta m_{VV'}=0.04\,m_\Phi$
    (i.e., $30\ {\rm GeV}$) for the $\gamma\gamma$ final state, and
    $\Delta m_{VV'}=0.12\,m_\Phi$ (i.e., $90\ {\rm GeV}$) for others.
    Here, we take $\sqrt{s_{ee}}=945\ {\rm GeV}$, $\sigma
    (pp\rightarrow \Phi\rightarrow\gamma\gamma; \sqrt{s_{pp}}=13\ {\rm
      TeV})=10\ {\rm fb}$, $\cos\theta_{\rm cut} = 0.9$, and $L_{ee} =
    1\ {\rm ab}^{-1}$.}
  \label{fig:etamin}
\end{figure}

Because the cross sections for the processes
$\gamma\gamma\rightarrow\Phi\rightarrow\gamma\gamma$, $\gamma Z$,
$ZZ$, and $W^+ W^-$ are proportional to $\eta$ with the present
parameterization, each mode is expected to be observed at the
photon-photon collider if $\eta$ is large enough.  In order to
estimate the minimal value of $\eta$ for the observation of each mode,
we calculate the following quantity:
\begin{align}
  S_{VV'} / \sqrt{N_{VV'}^{\rm (BG)}} \equiv
  \frac{L_{ee} \sigma (\gamma\gamma\rightarrow \Phi\rightarrow VV') \cos\theta_{\rm cut}}
  {\sqrt{N_{VV'}^{\rm (BG)}}}.
\end{align}
For the processes $\gamma\gamma\rightarrow\Phi\rightarrow\gamma Z$,
$ZZ$, and $W^+ W^-$, the uncertainties in the measurement of $m_{VV'}$
are expected to be dominated by jet energy resolution, concentrating
on the hadronic decay modes of weak bosons.  Thus, in the narrow width
case, $\frac{1}{2}\Delta m_{VV'}$ is taken to be twice the detector
resolution; we adopt $\Delta m_{VV'}=0.12\,m_\Phi$ for these final
states. For $\gamma\gamma\rightarrow\Phi\rightarrow \gamma\gamma$, we
take $\Delta m_{VV'}=0.04\,m_\Phi$, assuming a good resolution of the
electromagnetic calorimeter.  Then, we determine the minimal value of
$\eta$ which realizes $S_{VV'} / \sqrt{N_{VV'}^{\rm (BG)}}>5$.  The
result is shown in Fig.\ \ref{fig:etamin}.  We note that, if the
detector resolution (in particular, for hadronic objects) is not good
enough, hadronic decay of $W^\pm$ may be mis-identified as that of
$Z$ boson.  Then, the standard-model $W^+W^-$ production process may
also contribute to the background of the signal with $ZZ$ final state
with the hadronic decays of $Z$ bosons.  Such a problem may be avoided
by requiring that at least one of the $Z$ bosons decays into $e^+e^-$
or $\mu^+\mu^-$; if we impose such a requirement, the minimal value of
$\eta$ for $S_{ZZ} / \sqrt{N_{ZZ}^{\rm (BG)}}>5$ shown in
Fig.\ \ref{fig:etamin} should be divided by $\sim 0.36$.

We can see that, if $\eta\gtrsim O(10^{-2})$, cross sections for
several modes with electroweak gauge boson final states may be
measured at the photon-photon collider.  With measuring the cross
sections of two (or more) of such final states, the ratio
$\Lambda_1/\Lambda_2$ can be determined up to two-fold ambiguity.  In
addition, combined with the information about $\sigma
(\gamma\gamma\rightarrow\Phi\rightarrow gg)$ (or about $\sigma
(pp\rightarrow\Phi\rightarrow\gamma\gamma)$ at the LHC), the quantity
$\eta$ can be determined.  (The determination of $\eta$ may be
challenging only with the LHC experiment unless the cross section for
the process $pp\rightarrow\Phi\rightarrow gg$ is measurable at the
LHC.)  Thus, we will understand the relative size of $\Lambda_i$'s,
from which we will acquire information about the physics behind the
higher-dimensional operators given in Eq.\ \eqref{Lvarphi} or
\eqref{Lphi}.  We also note that, if a large amount of the event
sample of the process $\sigma (\gamma\gamma\rightarrow\Phi\rightarrow
ZZ)$ is obtained, CP property of $\Phi$ may be obtained from the
angular correlation of the decay products of $Z$ bosons.

So far, we have assumed that the di-photon events observed at the LHC
is due to the scalar-boson production via the gluon-gluon fusion.
Another possibility is the production of the scalar boson due to
photon-photon collision.  In such a case, the cross section for the
scalar-boson production at the LHC is estimated as
\cite{Csaki:2016raa}
\begin{align}
  \sigma_{\gamma\gamma}
  (pp\rightarrow \Phi\rightarrow\gamma\gamma; \sqrt{s_{pp}
  }=13\ {\rm TeV})
  \simeq 10.8\ {\rm pb} \times
  \left[ \frac{\Gamma_\Phi}{45\ {\rm GeV}} \right]
  {\rm Br}^2 (\Phi\rightarrow\gamma\gamma).
\end{align}
Then, for the narrow width case, we obtain
\begin{align}
  \sigma
  (\gamma\gamma\rightarrow \Phi\rightarrow\gamma\gamma; 
  \sqrt{s_{ee}}=945\ {\rm GeV})
  \simeq 4.7\ {\rm pb} \times
  \left[
    \frac{\sigma_{\gamma\gamma}
  (pp\rightarrow \Phi\rightarrow\gamma\gamma;
  \sqrt{s_{pp}}=13\ {\rm TeV})}{10\ {\rm fb}}
  \right].
  \label{sig_gammagamma}
\end{align}
(The cross section for other final state $F$ can be obtained by
multiplying Eq.\ \eqref{sig_gammagamma} by ${\rm Br}(\Phi\rightarrow
F)/{\rm Br}(\Phi\rightarrow\gamma\gamma)$.)  Thus, the cross section
for the $\gamma\gamma$ final state is orders of magnitude larger than
that of the background (see Table \ref{table:bgVV}), and hence the process
$\gamma\gamma\rightarrow \Phi\rightarrow\gamma\gamma$ will be observed
by the photon-photon collider.  In addition, the cross sections for the
$\gamma Z$ and $ZZ$ final states are typically of the same order of
that for the $\gamma\gamma$ final state unless $\Lambda_{\gamma Z}^{-1}$
and $\Lambda_{ZZ}^{-1}$ are extremely small.  The cross section for
the $W^+W^-$ final state can be also sizable unless $\Lambda_2\gg
\Lambda_1$.  Thus, in this case, detailed studies of the di-photon
resonance production with the electroweak gauge boson final states are
expected.  The detectability of the process of
$\gamma\gamma\rightarrow \Phi\rightarrow gg$ depends on ${\rm
  Br}(\Phi\rightarrow gg)$; $S_{gg}/\sqrt{N_{\bar{q}q}^{\rm (BG)}}$
becomes larger than $5$ at $L_{ee} =
  1 {\rm ab}^{-1}$ when ${\rm Br}(\Phi\rightarrow gg)/{\rm
  Br}(\Phi\rightarrow\gamma\gamma) \gtrsim 2\times 10^{-4}$.

Finally, we comment on the production of $\Phi$ with the $e^+e^-$
collision.  With the interaction given in Eq.\ \eqref{Lvarphi} or
\eqref{Lphi}, $\Phi$ can be produced with the process of
$e^+e^-\rightarrow\Phi\gamma$ and $\Phi Z$.  The $\Phi$ production is
also possible via the vector-boson fusion processes of
$e^+e^-\rightarrow\Phi e^+e^-$ and $e^+e^-\rightarrow\Phi
\bar{\nu}_e\nu_e$.  We have estimated the cross section for these
processes, taking the center-of-mass energy of $1\ {\rm TeV}$.  For
$\eta=1$ and $-10\leq\Lambda_1/\Lambda_2\leq 10$, the cross sections
for these processes are typically $10^{-1}-10^{-2}\ {\rm fb}$.  Thus,
the cross sections at the photon-photon collider are a few orders of
magnitude larger.  The discussion about the detectability of $\Phi$ at
the $e^+e^-$ collider requires detailed study of the backgrounds,
which we leave for future study.

In summary, we have studied the prospect of investigating a scalar
boson $\Phi$ at the photon-photon collider, assuming that the recently
observed LHC di-photon excess is due to the scalar boson.  Assuming
also that the scalar-boson production at the LHC is dominated by the
gluon-gluon scattering and that $\sigma
(pp\rightarrow\Phi\rightarrow\gamma\gamma)\sim 10\ {\rm fb}$ at
$\sqrt{s_{pp}} = 13$ TeV, the total cross section for the scalar-boson
production at the photon-photon collider with $\sqrt{s_{ee}}\simeq
945\ {\rm GeV}$ can be as large as $\sim 170\ {\rm fb}$ (taking
$m_\Phi=750\ {\rm GeV}$).  We have also seen that various decay modes
of the scalar boson may be observed.  Thus, if the existence of the
new scalar boson is confirmed with the next run of the LHC, the
photon-photon collider can play an important role to study the
detailed properties of the scalar boson.

\vspace{5mm}

\noindent {\it Acknowledgment}: The work is supported by Grant-in-Aid
for Scientific research Nos.\ 23104008 and 26400239.

\end{document}